# Spin-orbit-torque MRAM: from uniaxial to unidirectional switching

Ming-Han Tsai,[1] Po-Hung Lin,[1] Kuo-Feng Huang,[1] Hsiu-Hau Lin,[2] and Chih-Huang Lai[1]*

[1]Department of Materials Science and Engineering, National Tsing Hua University, Hsinchu, 300, Taiwan

[2]Department of Physics, National Tsing Hua University, Hsinchu, 300, Taiwan

*Correspondence and requests for materials should be addressed to C.-H. L. (chlai@mx.nthu.edu.tw)

**Abstract**

With ultra-fast writing capacity and high reliability, the spin-orbit torque is regarded as a promising alternative to fabricate next-generation magnetic random access memory. However, the three-terminal setup can be challenging when scaling down the cell size. In particular, the thermal stability is an important issue. Here we demonstrate that the current-pulse-induced perpendicular exchange bias can significantly relieve the concern of thermal stability. The switching of the exchange bias direction is induced by the spin-orbit torque when passing current pulses through the Pt/Co system with an inserted IrMn antiferromagnetic layer. Manipulating the current-pulse-induced exchange bias, spin-orbit-torque switching at zero field between states with unidirectional anisotropy is achieved and the thermal agitation of the magnetic moment is strongly suppressed. The spin-orbit torque mechanism provides an innovative method to generate and to control the exchange bias by electrical means, which enables us to realize the new switching mechanism of highly stable perpendicular memory cells.

Due to its excellent retention and speed[1], magnetic random access memory (MRAM) is a promising candidate to replace the traditional memories. To integrate with the logic architectures at reduced costs[2, 3, 4, 5, 6, 7], it is desirable to manipulate the magnetization with electrical means. In recent years, spin-transfer-torque (STT) MRAM attracts lots of attentions for its non-volatility, scalability and quick access time. Despite of these superior properties, the high operating current density is energy consuming and the common read/write path posts a challenging problem for optimization. In contrast with the STT MRAM, the spin-orbit-torque (SOT) provides a unique setup to switch the magnetic layers without passing current through the insulating tunneling barrier and opens up a new window for designing spintronic devices.[4, 5, 7, 8] Overcoming the shortcoming in the STT MRAM, the split of read and write paths in the SOT MRAM is able to achieve better endurance and less disturbed read errors.[9, 10, 11, 12]

The SOT in the heterostructure of a ferromagnetic (FM) layer and an adjacent non-magnetic (NM) layer arises from the strong spin-orbit interaction in the non-magnetic layer and/or the asymmetrical interfacial structure.[5, 13, 14] Applying an in-plane electrical current into the heterostructure, the magnetization of FM layer with perpendicular anisotropy can be manipulated by damping-like and field-like torques.[13, 14, 15] To achieve deterministic magnetic switching[4, 15], an additional external magnetic field along the current direction is usually necessary. This issue complicates the techniques for practical applications. Some resolutions have been proposed by introducing the lateral asymmetry in structure[16, 17] or inserting an antiferromagnet (AFM) layer[18, 19].

Another issue of the SOT MRAM is the longstanding challenge of maintaining both perpendicular anisotropy and thermal stability upon shrinking the size of devices. For instance, when scaling down the CoFeB-based magnetic tunnel junction (MTJ), the limited volume of FM layer seems to be an intrinsic obstacle for thermal stability[20, 21]. To improve thermal stability, an extra AFM layer adjacent to the free layer has been proposed for the thermally assisted STT-MRAM structure.[22, 23] Both the reference layer and the free layer are pinned by the AFM layers but with different blocking temperatures ($T_B$). The current injected into the junction heats the MTJ and the spin transfer torque defines the direction of the magnetization in the free layer. However, the heating temperature needs to surpass $T_B$ (~200 °C) of the free layer. In addition, the magnetization in this thermally assisted MTJ is in-plane.

The thermal stability issue can be solved differently in SOT MRAM. We investigate the NM/FM/AFM tri-layer system – the spin current is generated from NM layer and passes through the FM layer and even the FM/AFM interface. Because the spin-orbit torque induced by the current pulse is intrinsically non-equilibrium, not only the magnetization of FM layer can be reversed, but the uncompensated spins of the AFM at the interface are also flipped. In consequence, we can control the direction of exchange bias (EB) in the perpendicular direction without going through any annealing process.

By applying current pulses in opposite directions, the perpendicular EB switches accordingly as seen in the shifted hysteresis loops and creates the unidirectional anisotropy (along perpendicular direction) at zero field. The unidirectional anisotropy breaks the degeneracy between two minima along the easy axis and lifts one of them into the metastable state. Because there is only one true minimum left at zero field, the thermal stability is greatly enhanced. The EB switching and the unidirectional anisotropy significantly enhance the thermal stability and provide a promising resolution for future MRAM devices. Moreover, our findings point out the possibility to modify the interfacial properties between FM and AFM layers by current-pulse manipulation through the spin-orbit interactions. As will be elaborated later, one can make use of the current-pulse-induced EB to achieve zero-field switching as well, which removes another obstacle of SOT-MRAM for real applications.

**Result**

**Perpendicular exchange bias.** Here we demonstrate the perpendicular exchange bias in the as-deposited Ta (2.5)/Pt (2)/Co (1.2)/IrMn ($t_{\mathrm{IrMn}}$)/Pt (4)/Ta (2.5) stacks (in nm) with $t_{\mathrm{IrMn}} = 2, 4, 6, 8$ and $10$ nm (see Figure 1). The out-of-plane magnetic properties of samples were verified by the vibration sample magnetometer (VSM) and perpendicular easy axis is attained for all samples. As shown in Figure 1(c), perpendicular EB becomes apparent for samples with $t_{\mathrm{IrMn}} = 6, 8, 10$ nm, exhibiting dual loops with both positive and negative shifts. The schematic diagram of exchange coupling between Co and IrMn layers for the as-deposited sample with $t_{\mathrm{IrMn}} \geqq 6$ nm is shown in Figure 1(b). The dual-loop phenomenon is attributed to the Co domains with opposite orientations in the as-deposited samples. When $t_{\mathrm{IrMn}} \geqq 6$ nm, the AFM grain volume is large enough to hold the interfacial spins accordingly, creating the observed EB with both positive and negative shifts.[24, 25, 26, 27] Similar behaviors are also observed in FeMn-based exchange-bias systems with in-plane anisotropy.[28,29]

**Current-pulse-induced EB switching**. All the specimens are patterned from the as-deposited films into 5-$\mu$m-wide single wire with current pulses running along the $x$ axis as shown in Figure 2(a). The rise time and duration of the current pulses are 8.3 ns and 10 $\mu$s respectively as shown in Figure 2(b). Comparing with the rise time, a relatively longer fall time (1 $\mu$s) of the current pulse is chosen to achieve robust binary switch by the spin-orbit torque mechanism.[4, 30,31]

We applied the current pulses through the single wire ($t_{\text{IrMn}}$ = 8 nm) in the presence of an in-plane magnetic field ($H_{\text{x}}$ = 300 Oe) and measured the SOT switching by the focused polar magneto-optical Kerr effect (FMOKE). As shown in Figure 2(c), when the current density surpasses the threshold $J_c$ , the magnetization exhibits a sharp transition due to the spin-orbit torque. To further examine the magnetic property after the SOT switch, as shown in Figure 2(d)(e), we measured the out-of-plane hysteresis loop (no current flowing now) and found the EB shifts along with the SOT switching. That is to say, in the Pt/Co/IrMn tri-layer structure, the transition not only flips the magnetization but also changes the EB accordingly. It is rather remarkable that, with the insertion of the IrMn layer, the SOT switching between states with unidirectional anisotropy is achieved. In contrast with the SOT switching between degenerate states with uniaxial anisotropy, there is only one true minimum in the free-energy landscape with unidirectional anisotropy so the thermal stability is greatly enhanced. The samples with $t_{\text{IrMn}}$ = 6 nm and 10 nm display similar switching behavior, as shown in Supplementary S1.

**Joule heating.** It is known that EB can be established by annealing or deposition in magnetic field.[32,33,34,35] In the presence of the magnetic field, the FM spins line up with the field. The interfacial AFM spins then align ferromagnetically with the FM layer after the field-cooling process. In order to change the direction of the AFM spins adjacent to the FM layer, it usually requires temperature higher than the blocking temperature ($T_B$). Is it possible that, when current pulses pass through the wire, Joule heating boosts the local temperature exceeding $T_B$ of the AFM layer? If this occurs, the orientation of interfacial spins and magnetic moments can be re-defined.[22, 23]

However, our experimental findings indicate that Joule heating does not play the dominant role. First of all, we extract the blocking temperature $T_B$ from the shifts of the hysteresis loops (see Supplementary S2 for details). Compared with other IrMn-based Co/Pt multilayer systems, the extracted blocking temperatures for our samples are reasonable and follow the right trend with increasing $t_{IrMn}$.[36, 37] The local temperature due to Joule heating is estimated from the temperature-dependent resistance. The elaborated results can be found in Supplementary S2. The highest local temperature due to Joule heating for the sample with $t_{IrMn} = 8$ nm is about 90 ℃, much lower than the blocking temperature. Thus, the thermal fluctuations associated with Joule heating is not sufficient to explain the observed EB shifts induced by current pulses.

**SOT switching mechanism.** To understand the current-pulse-induced EB, we shall go back to its origin at microscopic scale. The magnitude of the EB can be modified by the electrical current due to either spin-transfer torque effect in FM or current-induced torques in AFM.[38] However, the EB set by the current pulse in a SOT MRAM follows a different mechanism. First of all, we conduct the lock-in measurement[39] (see Supplementary S3 for details) to study the SOT quantitatively and find that the spin Hall effect dominates here.[40] The dynamics of the magnetization under the influence of SOT is well captured by the generalized Landau-Lifshitz-Gilbert theory (see Supplementary S4 for details) developed in our recent work.[31] During the SOT switching, the dynamics of the unit vector of the magnetization $\hat{m} = (\sin\theta\cos\phi, \sin\theta\sin\phi, \cos\theta)$ is shown schematically in Figure 3. When the current density is smaller than the threshold value,

$$J_c = \frac{e}{\hbar}\frac{M_s t_F}{\theta_{\mathrm{SH}}}(H_K - \sqrt{2}\,H_x), \tag{1}$$

the magnetization remains in the initial direction, where $H_K$ is the effective anisotropy, $M_s$ is saturation magnetization, $t_F$ is thickness of the FM layer, $H_x$ is the applied external field along current direction and $\theta_{SH}$ is the spin-Hall angle. Above the current threshold $J > J_c$, its direction is determined by the torque balance between the longitudinal field and the spin-orbit interaction.

It turns out that the SOT drives the magnetization to a tilt angle $\theta_t$ above the plane as shown in Figure 3,

$$\theta_t = \sin^{-1}\left(\frac{H_x}{C_J}\right), \tag{2}$$

where $C_J = \frac{\hbar}{2e}\frac{\theta_{SH}}{M_s t_F}J$ denotes the strength of SOT. After the applied pulse is off, the SOT is gone and the magnetization relaxes to the stable minimum in the perpendicular direction, causing the observed binary transition of magnetization. It is worth emphasizing that the dynamics driven by the SOT is non-equilibrium in nature and cannot be understood as competition between two equilibrium free-energy minima along the perpendicular direction.

It is natural to expect that the spin current also causes non-equilibrium effects on the interfacial spins and thus affects the EB. Following similar analysis, the interfacial spins at FM/AFM are driven to some tilt angle $\theta_t$ by the SOT and then relax to the direction dictated by the ferromagnetic layer. Now the underlying mechanism is clear – the SOT switches the magnetic moment of FM layer and the uncompensated spins at the interface simultaneously and both relax into the same direction after the applied current pulse. Since the EB switches together with the magnetization, the SOT drives the transition between states with unidirectional anisotropy. Not only enhancing the thermal stability (only one minimum for the unidirectional anisotropy), it opens up the possibility to redefine the EB by the unconventional current-pulse approach.

**A tale of two EB's.** To deepen our understanding, we compared the EB generated by field-annealing and current-pulse approaches. First of all, the perpendicular EB can be achieved by field annealing at 200 ℃ for $t_{\mathrm{IrMn}}$ = 8 nm sample with magnetic field (~9000 Oe) along the $z$ direction. It is rather interesting to observe that, no matter how we set the EB by field annealing initially, its direction can be redefined after applying current pulses. But, the amplitudes of the bias field along the perpendicular direction are different: about 900 Oe by field annealing (see Supplementary S5 for details) while 600 Oe by SOT switching (as shown in Figure 2). The difference may result from distribution of exchange couplings among FM/AFM grains of various sizes. For larger AFM grains, more energy is needed to reverse the exchange bias direction[33, 41]. The field annealing may provide enough energy to define the majority of the interfacial spins. On the other hand, the SOT might not be sufficient to redefine the orientation of interfacial spins located on large AFM grains even after applying current pulses, rendering the overall exchange bias smaller.

The above behavior inspires us to design a fixed EB (set by field annealing) in the longitudinal direction while a dynamical EB (set by current pulses) in the perpendicular direction. Note that insertion of the AFM layer has been used as an effective in-plane field in the previous experiments, showing field-free SOT switching.[18, 19, 42, 43, 44] As shown in Figure 4, samples are annealed at 200°C for half an hour in magnetic field (~9000 Oe) along the $x$-axis. The in-plane EB of ~750 Oe sets in along the wire with details shown in Supplementary S6. Running current pulses through the wire will generate EB in the perpendicular direction but there is residual EB in the longitudinal direction. The residual EB along the current direction makes the field-free SOT switching possible. Our field-free SOT switching behaviors for the $t_{IrMn}$= 8 nm sample, shown in Figure 4, are comparable to other experiments[18, 19]. However, it is worth emphasizing that the field-free SOT switching (unidirectional) demonstrated here is different from previous experiments (uniaxial) because the current pulse also sets the direction of the perpendicular EB.

Figure 4 (b) represents the switching curves for the $+M_s$ initial state under various $H_x$. We can clearly observe the effect of the longitudinal field on the switching: when $H_x$ is reduced to 90 Oe, the magnetization reversal is not full.

However, even though the magnetization is not fully reversed, the SOT switching can be achieved at zero field ($H_x = 0$ Oe). Figure 4 (c) represents the switching curves for the initial state of $-M_s$ under various $H_x$. When $H_x = -90$ Oe, the magnetization remains the same during the sweep without SOT switching, inferring that the external field is fully compensated by the in-plane EB. With increased negative $H_x$, the magnetization reversal becomes more complete as expected.

The emergence of two types of EB's can be explained qualitatively in the following. The in-plane EB is first generated by field annealing. The applied current pulse, carrying SOT, reorients most of the interfacial spins of AFM grains and generates the dynamical EB in the perpendicular direction. But, a small amount of interfacial spins, possibly on large AFM grains, cannot be flipped by the current pulse and gives rise to the residual in-plane EB, serving as the effective longitudinal field crucial for zero-field SOT switching. Because field annealing and SOT switching are different mechanisms to generate EB, there is plenty of room to explore their subtle interplay and the resultant magnetic manipulations.

**The role of IrMn layer.** The insertion of the AFM layer brings up another surprise – the SOT switching is reversed with much lower current density threshold $J_c$. In the absence of IrMn layer, previous studies on Pt/Co/Pt trilayers show that the spin current is dominated by the top Pt layer.[40] However, with the insertion of AFM layer at the top interface, the SOT switching is reversed, implying the spin current is now dominated by the bottom Pt layer. To investigate the SOT switching systematically, it is helpful to study the SOT coefficient $\beta$ with different IrMn

thickness[45]:

$$\beta = \frac{2e}{\hbar} \frac{M_s t_F}{\theta_{SH}} = \frac{J_c}{\frac{H_K}{2} - \frac{H_x}{\sqrt{2}}}$$

The measured SOT coefficient $\beta$ is shown in Table 1 and plotted in Figure 5. With given $H_K$ and $H_x$, a smaller SOT coefficient $\beta$ implies the reduced current threshold and, thus, the SOT switching is easier to occur. If the direction of the total spin current is reversed, it will be reflected on the sign change of the SOT coefficient $\beta$. For thin IrMn insertion (1nm), both the current density threshold $J_c$ and the anisotropy field $H_K$ decreases but the SOT switching is not yet reversed. As the IrMn thickness increases to 2nm, the SOT switching is reversed. As shown in Figure 5, the SOT coefficient changes sign for larger IrMn thickness (2, 4, 6, 8, 10 nm). It was reported that IrMn possesses a finite spin Hall angle with the same sign as Pt but smaller in amplitude.[46, 47] Thus, simply adding contributions from the top Pt layer and the IrMn layer does not explain these experimental findings. It is plausible that the reduction in the current density threshold $J_c$ with increasing IrMn layer thickness may result from gradual blocking of spin current from the top Pt layer. Although we cannot completely neglect the spin current from the IrMn layer, we believe that our results reveal that the bottom Pt layer is the dominant spin current source for $t_{IrMn} > 1$ nm.

Since the dominant source of the spin current is now from the bottom Pt layer, the current density threshold $J_c$ shall change accordingly. Surprisingly, the efficiency of the net spin current is better with a lower current density threshold – the measured $J_c$ is smaller than half of that without the AFM insertion. The weakened perpendicular anisotropy from the intermixing of Co/IrMn alone cannot explain this strong reduction because the SOT coefficient $\beta$ is also reduced. In addition to the spin-current blocking by the IrMn insertion, the interfacial property of Co/IrMn may also play a crucial role in explaining the strong reduction of $J_c$. The effects of interfacial property on the SOT switching have been discussed in different systems.[48, 49] It is encouraging that one can reduce the SOT current density by interfacial engineering but the details need to be further investigated.

**Discussion**

To further investigate the phenomenon of the interface between Co and IrMn layer, we insert Cu or Ta in our stacks of structure at the top interface of Co (see Supplementary S7 for details). Because Cu and Ta are non-magnetic materials, they tend to lower the EB field as spacers. For Ta dusting layer (0.3 nm), the reduction of EB field is strong, consistent with the previous study reporting that Ta atoms may screen exchange coupling and form extended defects involving nearby atomic sites.[50] For Cu dusting layer, the effect on EB is much weaker. It is important to mention that, even though the EB field is reduced, the SOT coefficient $\beta$ becomes smaller showing better SOT switching efficiency. Although the EB field is reduced, the EB switching mechanism by SOT persists with the intervention of the dusting layer. The dusting layer may affect the strength of exchange coupling between Co and IrMn spins as well as the interfacial properties. Because SOT switching and current-pulse-induced EB sensitively depend on the interfacial properties, we have many knobs to optimize the desired magnetic properties through interface engineering.

To exclude the possibility that the EB switching behavior is caused by energy accumulation after each pulses, we apply a single pulse which was higher than the threshold value ( $J_c = \pm 3.38 \times 10^7 \text{A/cm}^2$ ) through the wire and measure the difference of hysteresis loops. As we expected, the variation of hysteresis loop is equivalent to the result shown in Figure 2, which confirms that the binary switching is not caused by the energy accumulation of the electric pulses. Furthermore, by sending in equal-magnitude pulses but with opposite directions in sequence, the binary switch is robust in the read-write measurement shown in Figure 6. The background noise arises from the poor reflection of polarized light in the structure for FMOKE detection and does not hinder the obvious binary switching behavior.

In conclusion, SOT is a promising candidate to manipulate magnetizations via electrical means. It is striking that applying current pulses not only achieve the SOT switching but also change the EB accordingly. In our samples, we show that the current-pulse-induced EB is larger than the coercivity field so that we can achieve only one single magnetization state at zero field. This finding provides an innovative method to redefine EB via electrical means and enhances the thermal stability tremendously. With further investigation and optimization along this direction, the current-pulse-induced EB shall have significant impacts on the SOT-MRAM and other spintronics devices.

**Methods**

**Sample preparation**

By ultra-high vacuum magnetron sputtering system, the Pt/Co/IrMn tri-layer structures were deposited on the thermally oxidized Si substrates without external magnetic field in the following order, sub./Ta(2.5 nm)/Pt(2 nm)/Co(1.2 nm)/IrMn($t$ nm)/Pt(4 nm)/Ta(2.5 nm). The bottom and top Ta were used for the adhesion and capping layer, respectively. We changed the thickness of IrMn layer to investigate the change of coercivity and the interfacial coupling between FM and AFM layers. By vibrating sample magnetometer (VSM), we confirm that all of the specimens show perpendicular magnetic anisotropy, as shown in Figure 1(c). After the examination of magnetic properties, these samples were patterned into Hall-bar (5 μm in width) or 5 μm single-line by photolithography and reactive-ion etching. For all the Hall-bar or single-wire samples, we further fabricated the Ta(10 nm)/Pt(100 nm) electrodes by photolithography, DC magnetron sputtering and lift-off process sequentially. For field-annealing processes, the samples were annealed in vacuum at 200℃ for half an hour under the in-plane or perpendicular magnetic field about 9000 Oe.

**Switching curve measurement**

We use the Focus-MOKE system to monitor the magnetization of the FM layer with a focused laser spot (5 μm in diameter) after each electric pulse with or without the magnetic field $H_x$. The electric pulses were applied by arbitrary wave form generator (Keysight 33509B), where the rise time and the fall time of the pulses can be modified, and monitored by the oscilloscope (Tektronix DPO05104B). The critical current density $J_c$ is calculated for the value that passes through the bottom Pt by assuming the film structure is a parallel resistor.

After electrical measurement, the out-of-plane hysteresis loops were inspected by the polar-mode of FMOKE.

**Author contributions**

M.H.T. and K.F.H designed the experiments and measurements. M.H.T and P.H.L made-up the samples. H.H.L and C.H.L constructed the explanation of phenomenon for the exchange-bias switching by SOT; M.H.T., H.H.L and C.H.L discussed and finished the manuscript. All authors provided the suggestions on the results and revised the manuscript.

**Competing financial interests:** The authors declare no competing financial interests.

**Figure legends:**

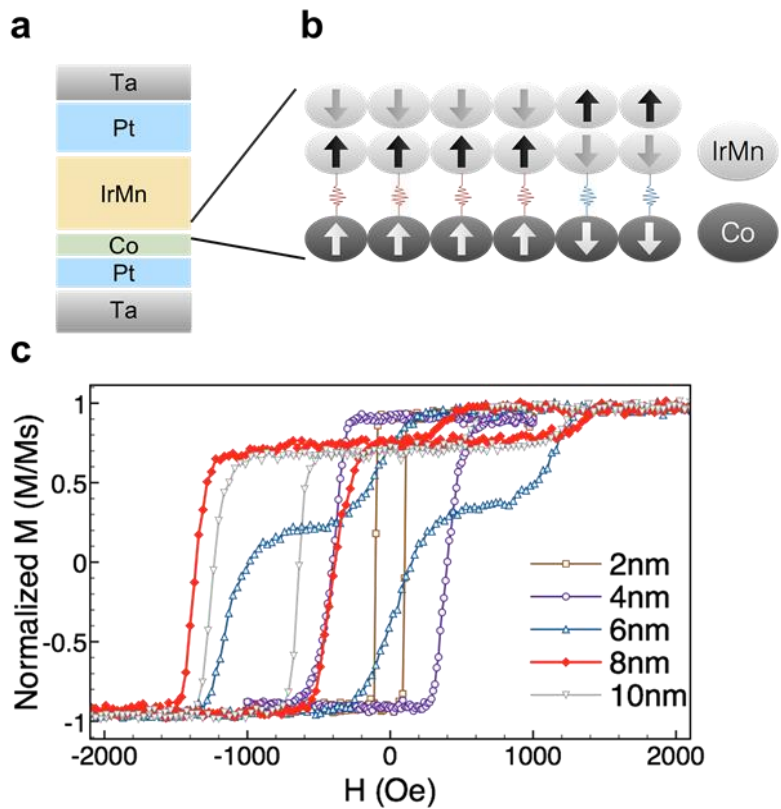

**Figure 1. Sample layout and magnetic properties.** (**a**) Cross-section of the sputter-deposited stack with perpendicular easy axis and EB. (**b**) Schematic plot for exchange coupling at the Co/IrMn interface in as-deposited samples. The red/blue lines represent interlayer couplings from different atomic moments, causing the magnetic moments of the Co layer pinned along different directions. (**c**) $M$-$H$ loops measured at the perpendicular direction of the film for $t_{\text{IrMn}}$ = 2, 4, 6, 8 and 10 nm.

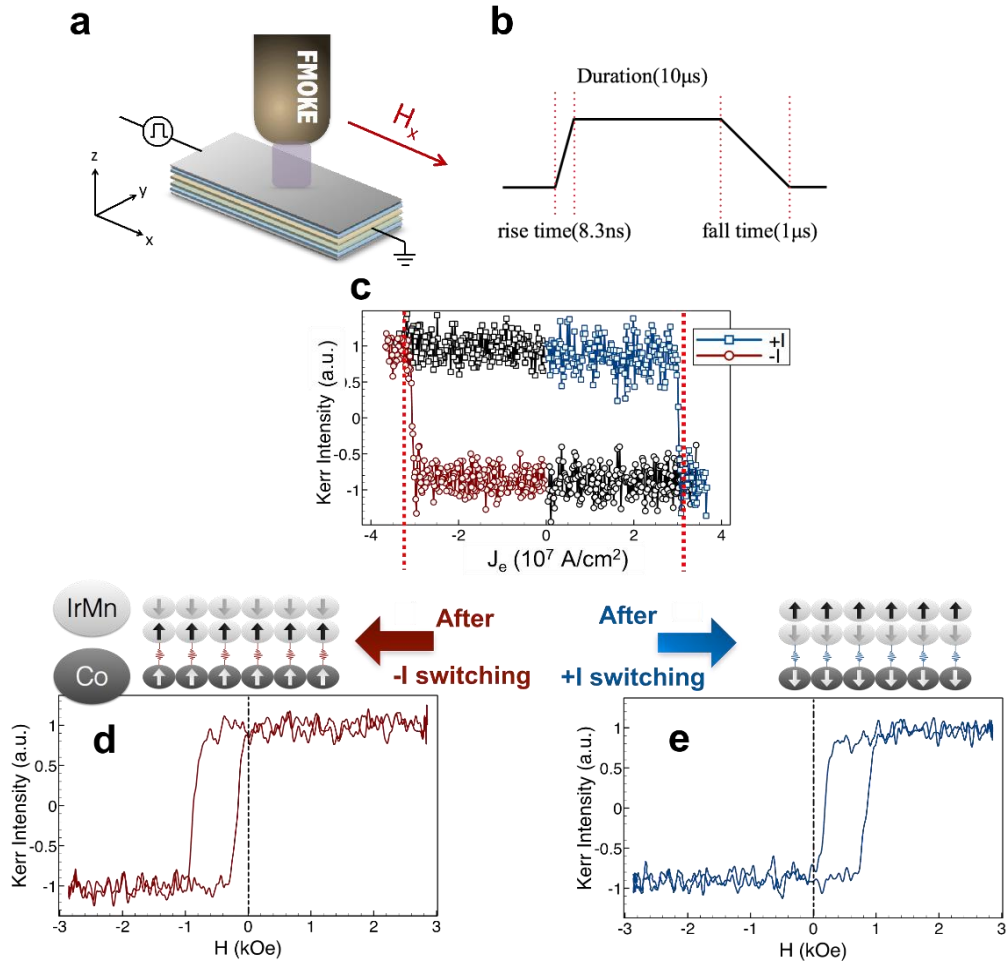

**Figure 2. SOT switching curves and the current-pulse-induced EB.** (**a**) Experimental setup for measuring the SOT switching curves by FMOKE. (**b**) The applied current pulse through the single-wire (5 μm by 10 μm), where duration = 10μs, rise time = 8.3ns and fall time = 1μs. (**c**) The SOT switching curve under the $H_x = 300$ Oe for $t_{IrMn}$ = 8 nm. (**d**)(**e**) The negative pulses (red line) reverse the magnetization from negative to positive, including the EB, and shift the easy axis loop to the left. The opposite behavior can be observed when we apply the positive pulses (blue line) through the wire.

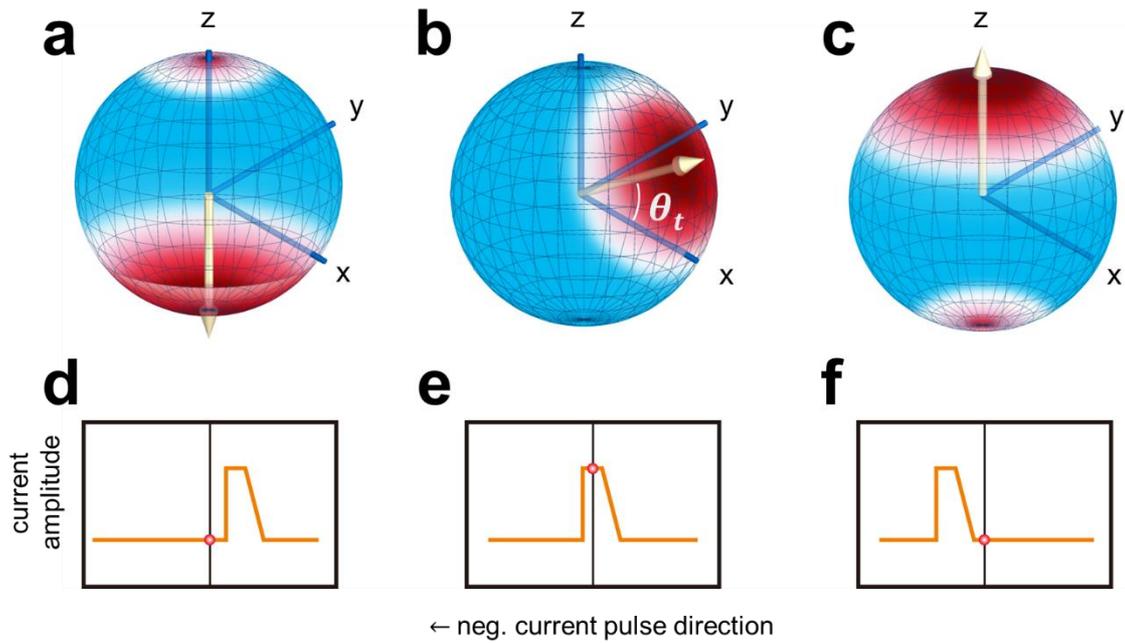

**Figure 3. Schematic diagram for SOT unidirectional switching.** (a-c) is the position of moment (represented by beige arrow) and its energy profile. The darkest red region denotes the minimum energy.

(d-f) represent the current pulse condition. (a, d) are the state before SOT switching (current is off). (b, e) are during the SOT switching. (current is on). (c, f) are after the SOT switching (current is off again). $\theta_t$ is the tilt angle.

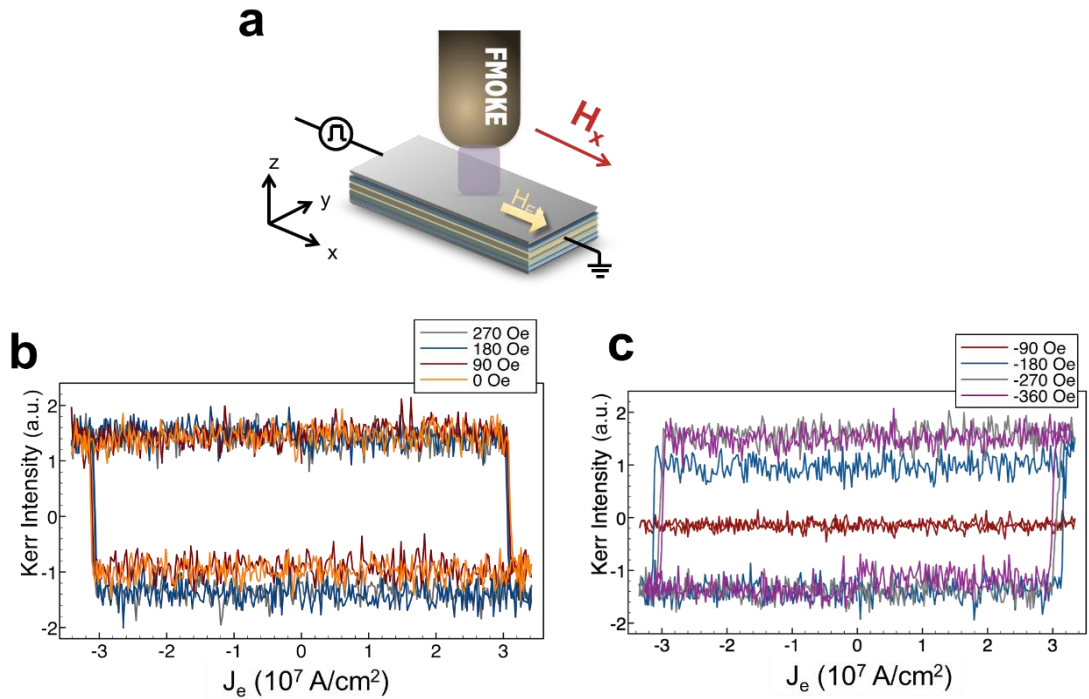

**Figure 4. Field-free SOT switching with different external magnetic field Hx.**
**(a)** Experimental setup for measuring the SOT switching curves by FMOKE without or with an external field Hx. The sample possesses an *in-plane* exchange bias. (b)(c) The magnetization reversal under the different external magnetic field for the samples with in-plane exchange bias. In (b), the external magnetic field Hx = 0, 90, 180, 270 Oe. In (c), the external magnetic field Hx = -90, -180, -270, -360 Oe. Here we can observe the deterministic switching in the absence of $H_x$

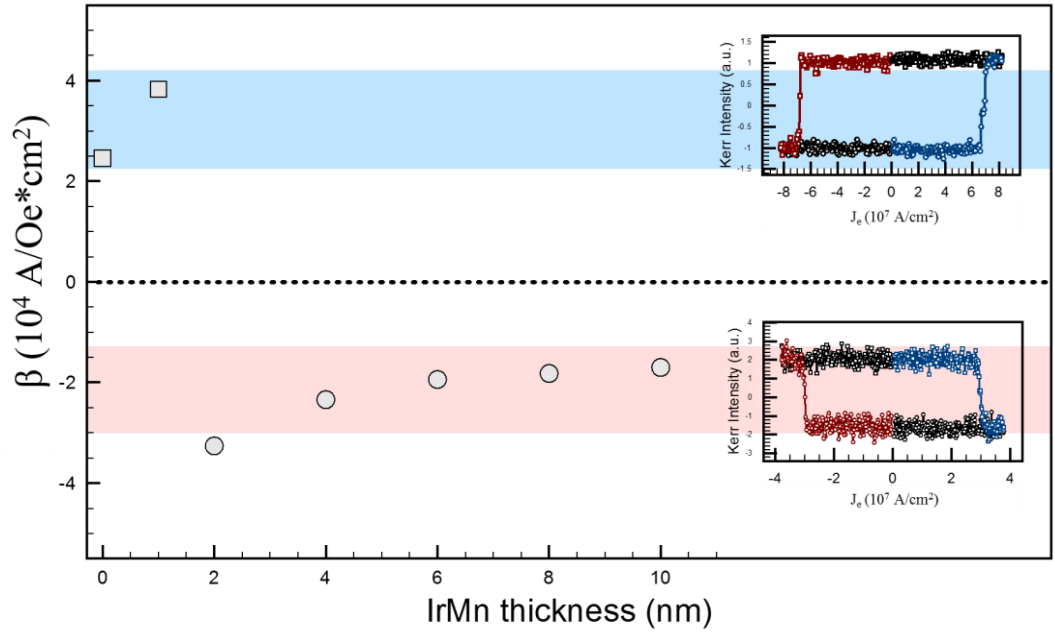

**Figure 5. Reverse SOT switching with larger IrMn thickness.** The SOT coefficient $\beta$ first increases with thin IrMn insertion (1 nm) and then changes sign with larger IrMn thickness (2, 4, 6, 8, 10 nm). The reverse SOT switching is evident via Kerr intensity measurements shown in the insets.

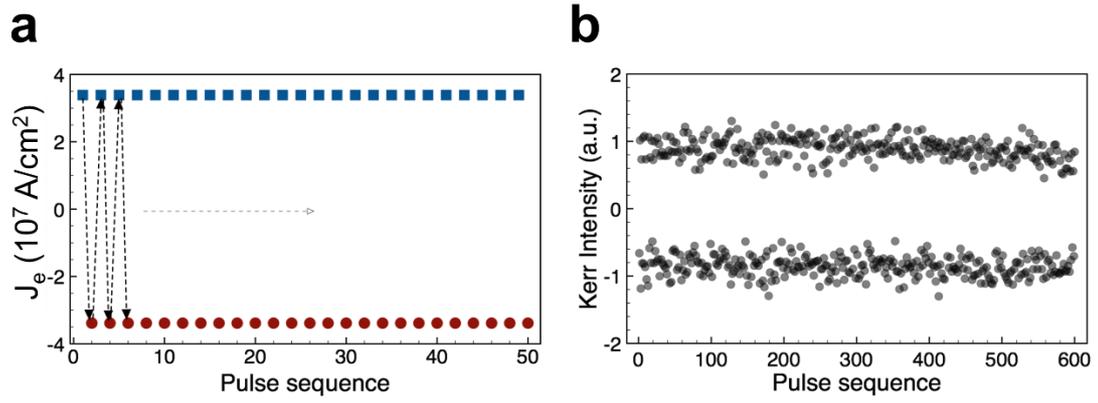

**Figure 6. Read-write test.** (**a**) Applied current pulse sequence with $H_x = 300$ Oe, (**b**) the resulting magnetization state after current pulses. The current polarity dependence of magnetization is clearly observed.

**Table 1: SOT coefficient $\beta$ for different IrMn thickness.**

| IrMn(nm) | 0nm | 1nm | 2nm | 4nm | 6nm | 8nm | 10nm |
|---|---|---|---|---|---|---|---|
| $J_c(10^7 A/cm^2)$ | 7.0 | 4.2 | 4.0 | 3.2 | 2.8 | 3.1 | 2.9 |
| $H_K$(Oe) | 6120 | 2620 | 2880 | 3160 | 3310 | 3840 | 3840 |
| $H_x$(Oe) | 300 | 300 | 300 | 300 | 300 | 300 | 300 |
| $\beta(10^4 A/(cm^2*Oe))$ | 2.46 | 3.83 | 3.26 | 2.34 | 1.94 | 1.82 | 1.70 |

**Supplementary Information**

**Spin-orbit-torque MRAM: from uniaxial to unidirectional switching**

Ming-Han Tsai,[1] Po-Hung Lin,[1] Kuo-Feng Huang,[1] Hsiu-Hau Lin,[2] and Chih-Huang Lai[1]*

[1]Department of Materials Science and Engineering, National Tsing Hua University, Hsinchu, 300, Taiwan

[2]Department of Physics, National Tsing Hua University, Hsinchu, 300, Taiwan

*Correspondence and requests for materials should be addressed to C.-H. L.

(chlai@mx.nthu.edu.tw)

**S1. SOT switching curves and the current-pulse-induced EB change for the samples with $t_{IrMn} = 6\ nm$ and $t_{IrMn} = 10\ nm$**

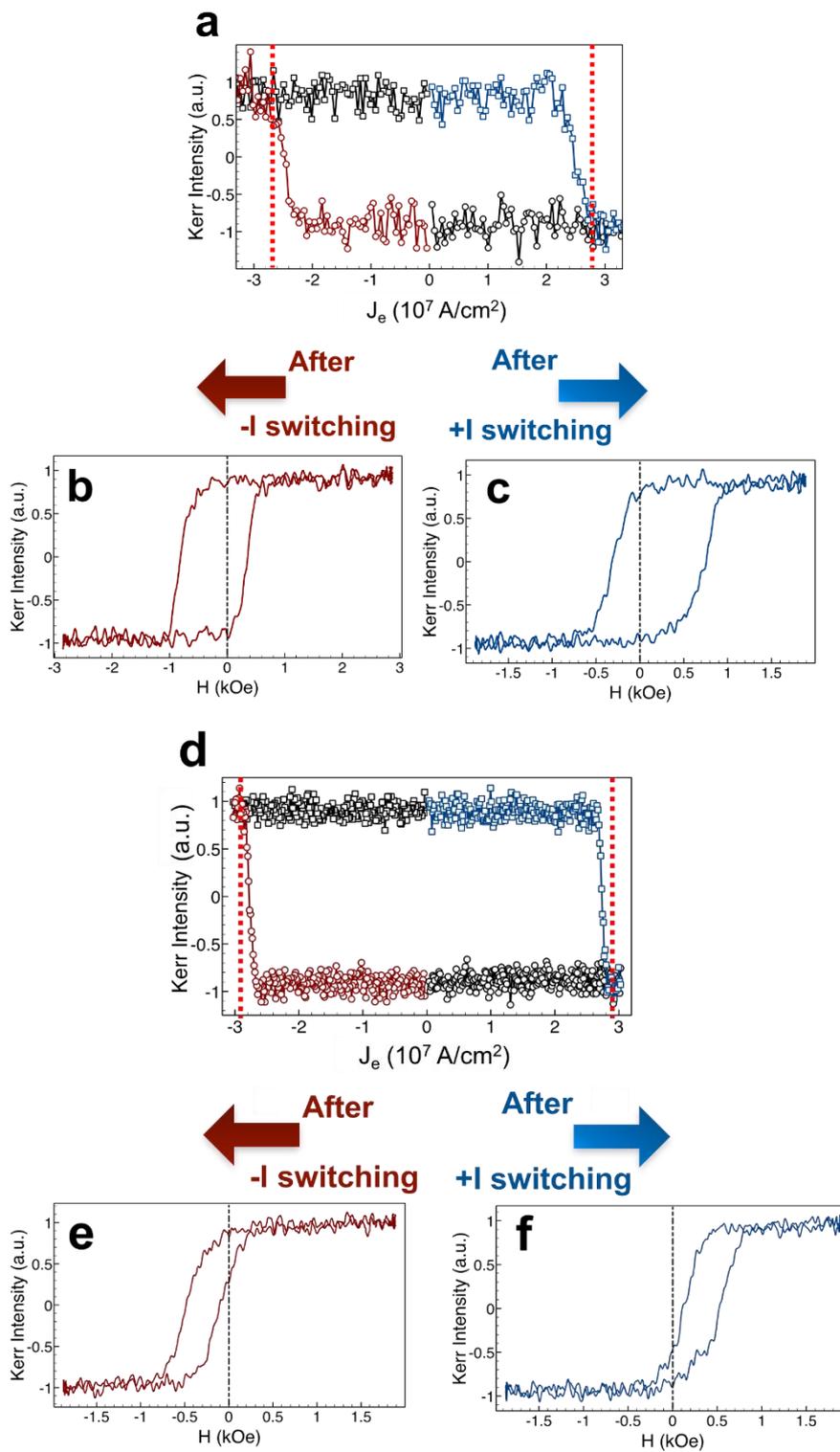

Figure S1. SOT switching curves and the current-pulse-induced EB change for the sample with (a-c) $t_{IrMn} = 6\ nm$ and (d-f) $t_{IrMn} = 10\ nm$.

The current pulse is applied through the single-wire (5 μm by 10 μm), where duration

$= 10\mu s$, rise time $= 8.3\text{ns}$ and fall time $= 1\mu s$. The SOT switching curves are measured under the $H_x = 300\,\text{Oe}$. The negative pulses (red line) reverse the magnetization from negative to positive, including the EB, and shift the easy axis loop to the left. The opposite behavior can be observed when we apply the positive pulses (blue line) through the wire.

**S2. Joule heating**

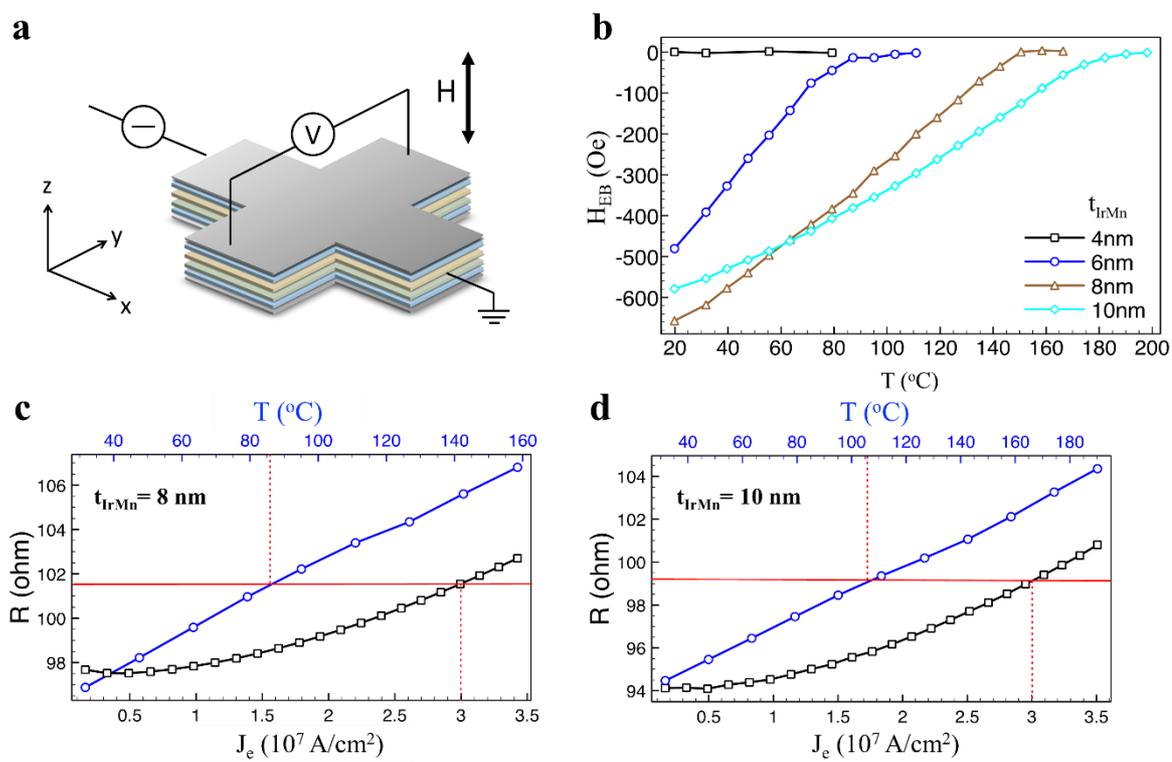

**Figure S2. Joule heating effect.** (**a**) Schematic setup for the Hall-bar measurement. (**b**) Blocking temperatures for the samples $t_{IrMn}$ = 4, 6, 8 and 10 nm, extracted from the temperature dependence of the hysteresis-loop shifts. The blocking temperatures $T_B$ for $t_{IrMn}$ = 8 and 10 nm are about 150°C and 180°C respectively. (**c**)-(**d**) Resistance for $t_{IrMn}$ = 8 and 10 nm at different temperatures and applied current densities. The red line represents the current-density threshold in SOT switching and the extracted temperature due to joule heating has not reached the blocking temperature.

With current pulses passing through the wire, the local temperature increases due to Joule heating. If the increased temperature exceeds the blocking temperature ($T_B$) of the AFM layer, it can re-define the orientation of interfacial spins and magnetic moments.[1, 2] It is crucial to check whether Joule heating plays the dominant role in the observed EB shifts. Here we use the Hall-bar structure (5 μm in width, Figure S2(a)) to estimate the exchange bias field ($H_{\mathrm{EB}}$) by anomalous Hall effect. First, we post anneal the specimens of different $t_{\mathrm{IrMn}}$ under a perpendicular magnetic field to create the out-of-plane EB. Afterwards, $H_{\mathrm{EB}}$ of the out-of-plane hysteresis loops at different temperatures are measured to extract the blocking temperature as shown in Figure S2(b). Compared with other IrMn-based Co/Pt multilayer systems, the extracted $T_{\mathrm{B}}$ in our samples are reasonable, following the right trend with increasing $t_{\mathrm{IrMn}}$.[3, 4]

In Figure S2(c), S2(d), sample resistance at different temperatures (blue line) and with different applied current densities (black line) are compared to estimate the local temperature due to Joule heating. For $t_{\mathrm{IrMn}}$ = 8 nm, the current density threshold $J_{\mathrm{c}} \cong 3 \times 10^7$ A/cm$^2$ corresponds to a local temperature less than 90 °C (red line), much lower than the blocking temperature around 150 °C. The local temperature due to Joule heating for $t_{\mathrm{IrMn}}$ = 10 nm is around 105 °C as shown in Figure S3(d), still lower than the blocking temperature around 180 °C . These results show that the observed EB shifts arise mainly from the SOT mechanism not the Joule heating in the samples.

## S3. Quantitative study of SOT by lock-in measurement

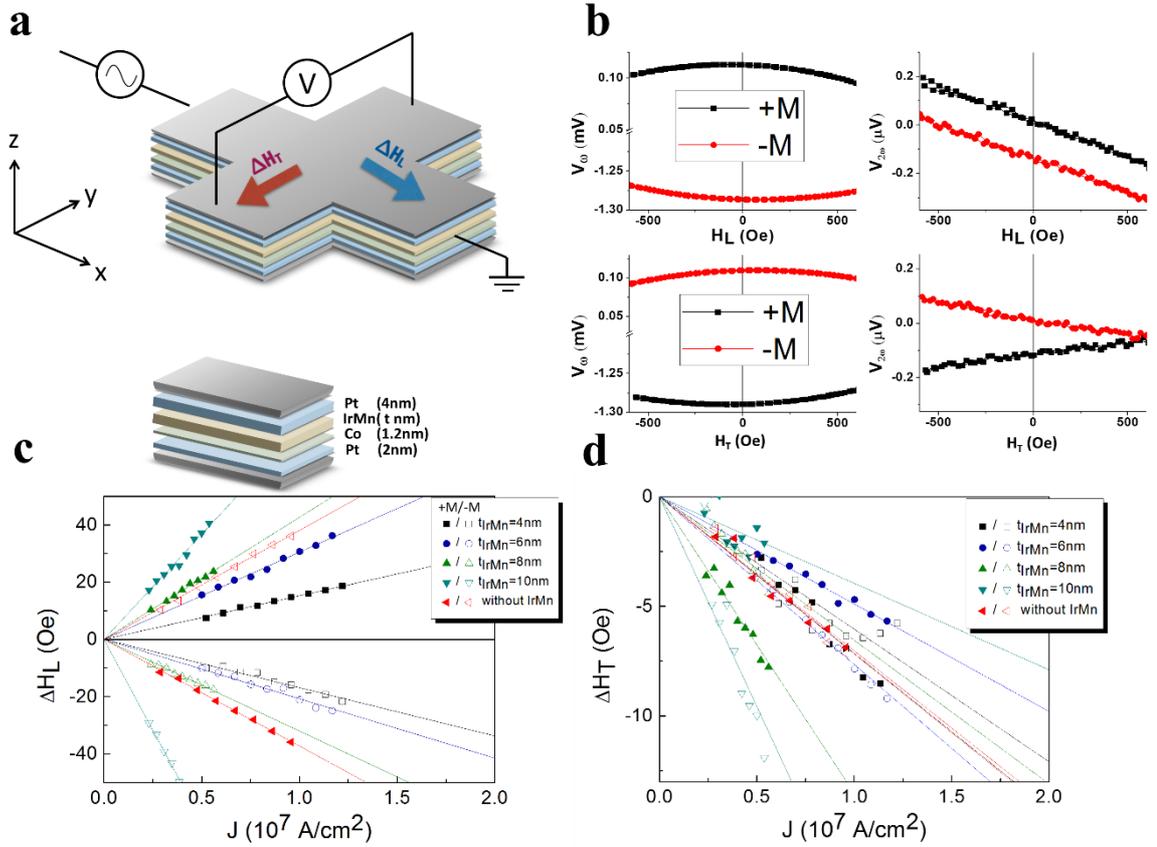

**Figure S3. Lock-in measurement to classify SOT.** (**a**) Schematic setup for the lock-in measurement for SOT effective field in longitudinal ($\Delta H_L$) and transverse ($\Delta H_T$) directions. (**b**) Typical in-phase first harmonic ($V_\omega$) and out-of-phase second harmonic ($V_{2\omega}$) Hall voltages versus magnetic field sweeping in longitudinal and transverse directions. (**c**) The longitudinal effective field $\Delta H_L$ and (**d**) the transverse effective field $\Delta H_T$ versus current density for samples with different $t_{IrMn}$ thickness and initial magnetization states (+M/-M).

In order to quantitatively estimate the spin-orbit torque, the lock-in experiment is conducted on the Hall-bar structure for acquiring the in-phase first harmonic ($V_\omega$) and the out-of-phase second harmonic ($V_{2\omega}$) Hall voltages. The direction of applied current ($x$ axis) and the direction of measuring voltage ($y$ axis) are defined as the longitudinal direction and the transverse direction respectively, as shown in Figure S3(a). We first apply a constant sinusoidal current to the wires and measure the Hall voltage, and then we sweep the in-plane field directed transverse or parallel to the current flow to obtain the transverse and the longitudinal components of the effective field vector. Figure S3(b) represents $V_\omega$ and $V_{2\omega}$ curves with external magnetic field sweeping along either longitudinal (ΔH$_L$) or transverse (ΔH$_T$) direction.

Both effective fields can be obtained by the following formula[5]:

$$\Delta H_{T(L)} = -2\frac{\partial V_{2w}}{\partial H_{T(L)}} \Big/ \frac{\partial^2 V_w}{\partial H_{T(L)}^2}$$

computed from the ratio between the slope of $V_{2w}$ and the curvature $V_w$ along the longitudinal or the transverse direction. The magnitudes and the signs of longitudinal (ΔH$_L$) and transverse (ΔH$_T$) effective fields versus current density in the samples are shown in Figure S3(c), S3(d). The amplitudes of ΔH$_L$ and ΔH$_T$ increase linearly with the current density. Note that the transverse direction corresponds to the Rashba interaction and/or spin Hall effect while the longitudinal direction corresponds to the spin Hall effect. We found that the amplitude of ΔH$_T$ was 4-10 times smaller than ΔH$_L$, hinting the spin Hall effect dominates here. The lock-in measurement provides a quantitative mean to classify the nature of the spin-orbit interactions. Note that planar Hall effect correction for the measured effective field is insignificant due to strong PMA of the Pt/Co.

In the multi-layer structure, there are multiple sources for spin currents such as top/bottom Pt layers[6-8] and IrMn layer[9,10]. Comparing the switching behavior studied in similar structures[7,11], the net spin current is provided by the bottom Pt layer, confirmed by the observed positive spin Hall angle. The switching behavior caused by the bottom Pt layer can also be verified by comparing with the sample without IrMn layer where the top Pt layer is the dominant source of spin current. The reverse SOT switching is evident in Figure S3(c) where the slope of $\Delta H_L$ changes sign when the IrMn layer is present.

**S4. Generalized Landau-Lifshitz-Gilbert equation**

Let us zoom into the details of magnetization dynamics when multiple domains are present. The generalized Landau-Lifshitz-Gilbert equation for the magnetization of each active magnetic cluster reads

$$\frac{\partial \hat{m}}{\partial t} = -\gamma \, \hat{m} \times \vec{H}_{\text{eff}} + \alpha \, \hat{m} \times \frac{\partial \hat{m}}{\partial t} + \gamma \, C_J \, \hat{m} \times (\hat{m} \times \hat{y}) \qquad (S1)$$

where $\hat{m} = (\sin\theta\cos\phi \, , \sin\theta\sin\phi \, , \cos\theta)$ is the unit vector of the magnetization, $\gamma$ represents the gyromagnetic ratio, $\alpha$ is the damping constant, $\vec{H}_{\text{eff}}$ is the effective magnetic field and $C_J = \frac{\hbar}{2e}\frac{\theta_{\text{SH}}}{M_s t_F} J$ denotes the strength of the toque from the spin-orbit interaction. The index to label different domains is suppressed for notation clarity. The magnetization quickly damps into the steady state ($\frac{\partial \hat{m}}{\partial t} = 0$) with direction determined by the torque balance between the longitudinal field and the spin-orbit interaction. For current density smaller than the critical value,

$$J_c = \frac{e}{\hbar}\frac{M_s t_F}{\theta_{\text{SH}}}\left(H_K - \sqrt{2}\,H_x\right) \qquad (S2)$$

the magnetization remains in the initial direction. The theory predicts that the critical current density $J_c$ changes linearly with the longitudinal field $H_x$ as observed in the experiment. For current densities above the threshold $J > J_c$, the torque balance gives rise to the following criterion,

$$H_x \sin\theta \sin\phi = C_J \, \cos\theta \, \sin\theta \sin\phi \qquad (S3)$$

For realistic parameter, the torque-balanced angle $\theta$ is close to $\pi/2$ (almost in the current-flowing plane). Thus, it is more convenient to introduced the tilt angle $\theta_t$ of the magnetization above the plane,

$$\theta_t = \frac{\pi}{2} - \theta = \sin^{-1}\left(\frac{H_x}{C_J}\right) \qquad (S4)$$

The tilt angle determines whether the magnetization relaxes to either up or down directions with specific probabilities and causes the SOT switching in experiment. Here we choose a relatively longer fall time so that the relaxation of the magnetization is guided to the unidirectional minimum with almost unity probability.[12] This is consistent with the robust binary transition observed in experiment.

**S5. Exchange bias induced by field annealing and SOT**

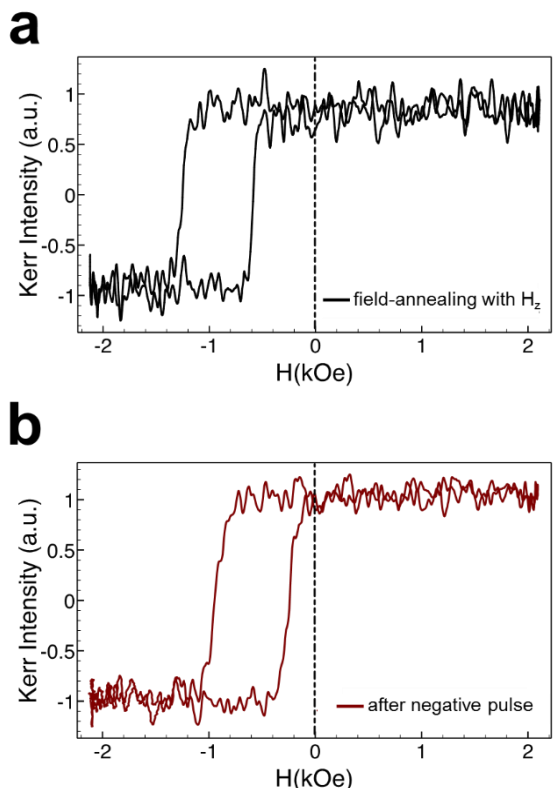

**Figure S5. The hysteresis loop for the single-wire sample (IrMn=8nm) with field-annealing and after the SOT switching.** (a) The hysteresis loop of the sample with the field-annealed along the z-direction. The $H_{EB}$ is about 900 Oe. (b) The hysteresis loop of the sample after the SOT switching with negative current pulse. The $H_{EB}$ is about 600 Oe.

The perpendicular EB can be achieved by field annealing at 200 ℃ for $t_{IrMn}$ = 8 nm sample with magnetic field (~9000 Oe) along the $z$ direction, shown in Figure S5 (a). The exchange bias can also be set by SOT, shown in Figure S5 (b). However, the amplitudes of the bias field along the perpendicular direction are different: about 900 Oe by field annealing while 600 Oe by SOT switching.

**S6. Evidence of the in-plane EB by field annealing along in-plane direction**

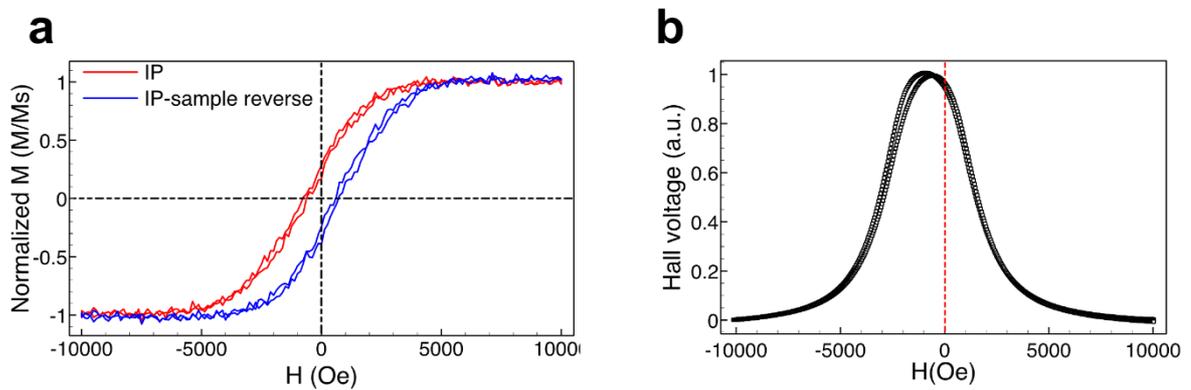

**a**

**b**

Figure S6. (a) The hysteresis loop measured by VSM for sheet film with in-plane field-annealing. The in-plane $H_{EB}$ is about 750 Oe. The blue curve is the measurement for rotating the sample in-plane by 180 degree to reconfirm the existence of in-plane exchange bias. (b) The AHE measurement for the hall cross device along the in-plane direction. The device with IrMn=8nm is field-annealed along the in-plane direction. The shift of in-plane AHE curves confirm the existence of in-plane exchange bias in the device.

## S7. Insertion of Cu and Ta dusting layer

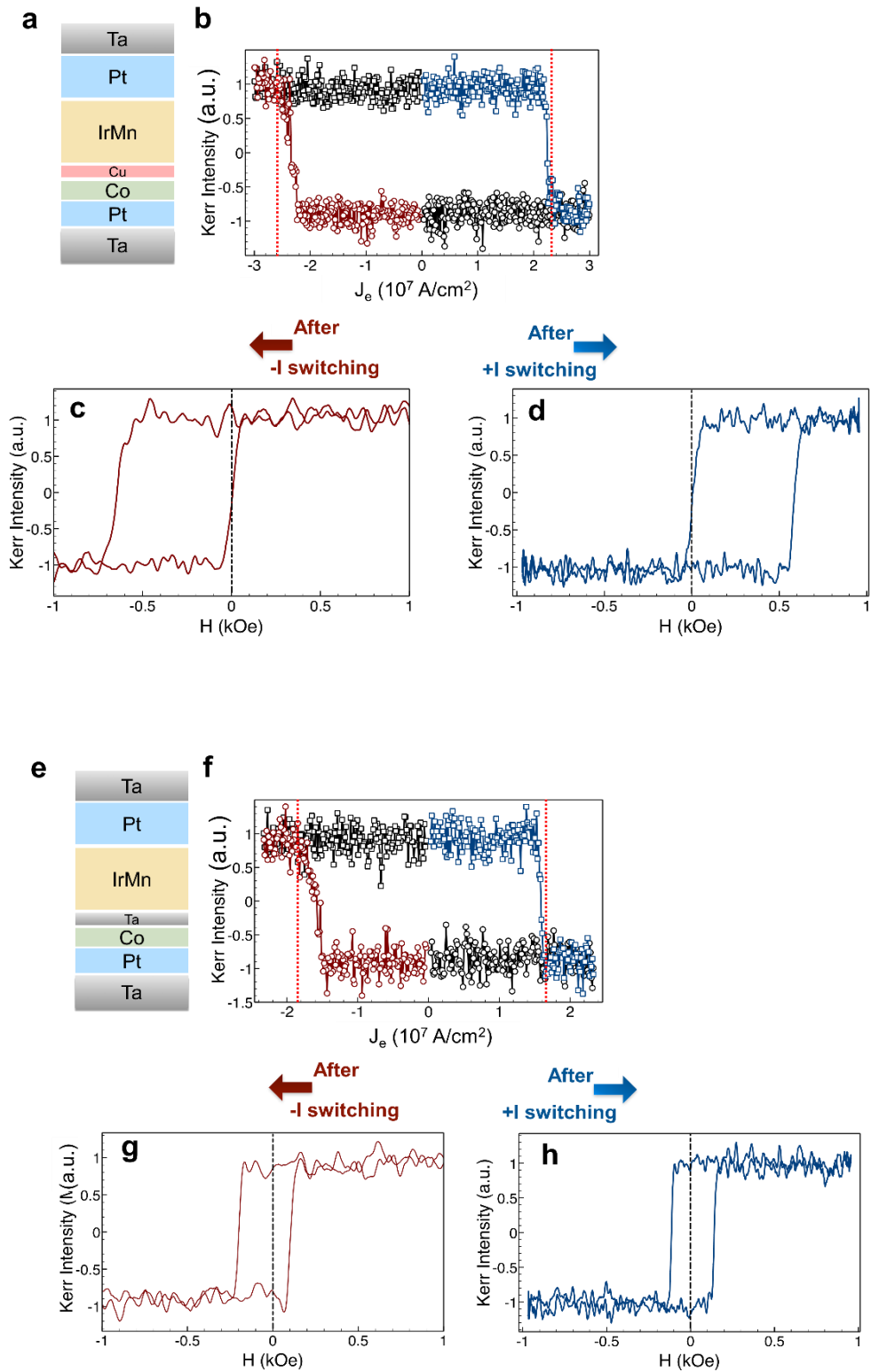

**Figure S7. SOT switching curves and the current-pulse-induced EB with Cu and Ta insertion.** (**a**)(**e**) Cross-section of the stack structure. (**b**)(**f**) The SOT switching curve with 0.3 nm Cu or Ta dusting layer ( $H_x = 300$ Oe, $t_{IrMn}$ = 8 nm). (**c**)(**d**)(**g**)(**h**) The negative pulses (red line) reverse the magnetization from negative to positive and both of EB and hysteresis loop shift to the left. The opposite behavior is observed when positive pulses are applied (blue line) through the wire. It is clear that different dusting layer insertion leads to different EB reduction.

For $t_{IrMn}$ = 8 nm sample with the insertion of Cu dusting layer (0.3 nm), the current-pulse-induced EB decreases, as shown in Figure S7(c), S7(d). Because Cu atoms are light and non-magnetic, the dusting layer presumably serves as spacer. The effect of the Ta dusting layer is more dramatic with almost vanishing EB field. It is reported that, in comparison with Cu, Ta has substantial influence on the grain size and crystal structure of IrMn[13,14]. In consequence, not only we observe the strong reduction of EB field but also the reduced coercivity field (Figure S7 (g), S7(h)). Furthermore, Ali *et al*[15] reported that Ta atoms may screen the exchange coupling between Co and IrMn layer involving nearby atomic sites by creating extended defects. Nonetheless, the enhancement of EB field by dusting layer with different materials such as Fe and Pt has also been studied. It should be emphasized that the interfacial property provides a tuning knob for current-pulse-induced EB.

The dusting layer has a strong impact on the current density threshold $J_c$. For the Cu dusting layer, the threshold decreases to $J_c = 2.4 \times 10^7 \mathrm{A/cm^2}$. Because the anisotropy field $H_K$ remains more or less the same, the SOT coefficient $\beta = 1.34 \times$ $10^4 \text{A}/(cm^2 \text{Oe})$. And for Ta dusting layer, the current density threshold drops to $J_c = 1.7 \times 10^7 \text{A}/\text{cm}^2$ and the SOT coefficient is $\beta = 1.64 \times 10^4 \text{A}/(cm^2 \text{ Oe})$ because of the drop on $H_K = 2500$ Oe. As explained in the main text, the dusting layer changes the interfacial property and lowers the SOT coefficient $\beta$ significantly. There is still plenty of room for optimization by interfacial engineering to make the best SOT device.

**Supplementary reference:**